\documentclass[preprint]{aastex}
\def\simlt{\lower.5ex\hbox{$\; \buildrel < \over \sim \;$}}
\def\simgt{\lower.5ex\hbox{$\; \buildrel > \over \sim \;$}}

\shortauthors{Corbelli et al.}
\shorttitle{Sharp HI edges at high z}

\begin{document}

\title{Sharp HI edges at high z: the gas distribution
from Damped Lyman-$\alpha$ to Lyman-limit absorption systems}

\author{Edvige Corbelli,}
\affil{Osservatorio Astrofisico di Arcetri, Largo E.Fermi, 5,
50125 - Italy}
\email{edvige@arcetri.astro.it}

\author{Edwin E. Salpeter}
\affil{Department of Astronomy, Cornell University, Ithaca,
NY 14850}
\email{salpeter@astrosun.tn.cornell.edu}

\and

\author{Rino Bandiera}
\affil{Osservatorio Astrofisico di Arcetri, Largo E.Fermi, 5,
50125 - Italy}
\email{bandiera@arcetri.astro.it}

\begin{abstract}

We derive
the distribution of neutral and ionized gas in high redshift clouds 
which are optically thick to hydrogen ionizing radiation, using published 
data on Lyman-limit and Damped Lyman-$\alpha$ absorption systems in the 
redshift range 1.75$\le z < $3.25. We assume that  
the distribution of the hydrogen total (HI+HII) column density in the
absorbers, $N_H$, follows a power law  $K N_{H}^{-\alpha}$, whereas
the observed HI column density distribution deviates from a pure power 
law as a result of ionization from a background radiation field.
We use an accurate radiative transfer code for computing the
rapidly varying ratio $N_{H}/N_{HI}$ as a function of 
$N_{H}$. Comparison of the models and observations gave
excellent fits with Maximum Likelihood solutions for
the exponent $\alpha$ and for $X$, the value of 
log($N_{H}/N_{HI}$) when the Lyman-limit optical depth
along the line of sight is $\tau_{LL}=1$. 
The slope of the total gas column density distribution  
with its relative 3$\sigma$ errors  is $\alpha=2.7^{+1.0}_{-0.7}$
and $X=2.75\pm0.35$. This value of $X$ is much lower than what would be 
obtained for a gaseous distribution in equilibrium under its own gravity. 
The ratio $\eta_0$ of dark matter to gas density
is however not well constrained since log($\eta_0)=1.1\pm 0.8$.
An extrapolation of our derived power law distribution towards systems of 
lower column density, the Lyman-$\alpha$ forest, tends to favour
models with log$(\eta_0) \simlt 1.1$ and $\alpha \sim 2.7$ to 3.3. 
With $\alpha$ appreciably larger than 2, Lyman-limit systems contain more gas
than Damped Lyman-$\alpha$ systems and Lyman-$\alpha$ forest clouds even more.
Estimates of the cosmological gas and dark matter density due to absorbers
of different column density at $z\sim 2.5$ are also given.

\end{abstract}


\section{Introduction}

Lyman limit systems (hereafter LLS) are detected as intergalactic clouds 
which absorb the   
quasar radiation at energies above the ionization edge of neutral 
atomic hydrogen. The optical depth to the continuum radiation at the
Lyman edge is $\tau_{LL}= N_{HI}/(1.6\times 10^{17} {\hbox{cm}}^{-2})$, 
where $N_{HI}$ is the HI column density along the line of sight. For
$\tau_{LL}\sim 1$, $N_{HI}$ can be measured accurately regardless of
the velocity structure of multiple components; for $\tau_{LL}\gg 1$ the
absorption can be detected, although $N_{HI}$ cannot be measured. We shall
not analyze in detail the Lyman-$\alpha$ forest systems, at much smaller 
$N_{HI}$, but will consider the whole range of column densities from the LLS
to the, mostly neutral, damped Lyman-$\alpha$ systems (hereafter DLS) with
much larger $N_{HI}$. We will then discuss a possible extrapolation of
our results towards the Lyman-$\alpha$ forest region.
There have been some suggestions that LLS, as well
as MgII absorbers with similar HI column densities, may be related to
outskirts of galaxies \citep{bb91,gar99,rt00}
while DLS are more likely to originate from
their inner regions \citep[and references therein]{pw98}. 
We shall show that, whether this suggestion is
verified or not, the ratio of the total to neutral hydrogen column density 
for typical LLS can be estimated assuming only that the total gas 
column density distribution $g(N_{H\perp})$ is a smoothly varying function 
from LLS to DLS range. In
particular we shall derive a value for $X$, the logarithm of $N_{H\perp}/
N_{HI\perp}$
for a line of sight value of $N_{HI}=1.6\times 10^{17} {\hbox{cm}}^{-2}$, 
at intermediate redshifts, namely $1.75\simlt z\simlt 3.25$. 
From $X$ we shall derive
$\alpha$, the exponent in the power law for $g$. 

If the dimensionless parameter $X$ is very large, the ionization fraction
changes rapidly and the value of $N_{HI}$ can increase from $\sim 10^{17}$
cm$^{-2}$ to $\sim 10^{19}$ cm$^{-2}$ over a  fairly small increase of
the total column density. Breaks have been found in the
distribution of $N_{HI}$ of absorbers from the Lyman-$\alpha$ forest to
the DLS \citep{pet93,sto96,sw00}. We shall show that this behavior of 
$N_{HI}$ is compatible 
with a single power law distribution for the total hydrogen column density,
once the change in ionization
fraction is taken into account. In fact it is the change in slope
of the $N_{HI}$ distribution from the LLS to DLS region
which enables us to derive a value for $X$.

A similar phenomenon is found in the outskirts of today's galaxies where a
sharp decline of the HI column density over a narrow range of the galaxy
radius occurs. One can explain this occurrence in term of an HI-HII 
transition zone which takes place when the column density gets sufficiently low
that the gas becomes optically thin to the extragalactic ionizing flux. 
For todays galaxies the known rotational velocity constrains the dark  
matter content which, together with the measured radial decline of the HI 
column density at the HI-HII transition zone, can be used to estimate   
the UV ionizing flux at $z=0$, otherwise unobservable. \citet{cor93}
have in fact modeled the sharp HI decline observed in M33 by
considering a total gas distribution compressed by the local dark
matter (inferred from the observed rotational velocities), and irradiated 
by the extragalactic UV and soft X-ray background. They found a best fit to 
the observed HI radial distribution in the outermost disk and to its sharp 
decline when the intensity of the background radiation at the Lyman-edge is
$6\times 10^{-23}$ ergs cm$^{-2}$ s$^{-1}$ sr$^{-1}$.
For LLS the situation is reversed since the 
metagalactic ionizing flux at high $z$ is better known than the dark matter 
content of absorbers: from $X$ we shall derive the volume gas density $n$,
which is larger than self-gravity alone can produce and therefore
gives information on the dark matter gravitational potential. 

Unfortunately the difficulty 
in measuring the residual flux of LLS at large optical depths limits the 
determination of the HI column density of the absorber above $10^{18}$ 
cm$^{-2}$. For this reason we are forced to 
introduce a new technique for treating large uncertainties in the $N_{HI}$ 
values. In the construction of a database for LLS we
include also the high column density systems,
known as Damped absorbers, whose column density is 
determined from the scatter of the Lyman-$\alpha$ radiation. Both
the database and the data treatment are described in detail in 
\citet[hereafter Paper II]{baco}. In section 2 of this paper we 
derive the number density of LLS and DLS from our database and in section 3
we describe radiative transfer processes and the gas stratification model
for the gas in LLS and DLS.
The  power law distribution for the total hydrogen column density 
and the H fractional ionizations which fit at best the data at 
$1.75\le z < 3.25$ in the Lyman-limit and Damped Lyman-$\alpha$ region will 
be derived in section 4. Having the H ionization
fractions for LLS, we can then compute the total gas content of 
LLS and DLS and estimate their dark matter potential.  
Also given in section 4 is an extrapolation of our fit towards absorption
systems of even lower HI column density.

\section {Number density of Lyman limit systems and Damped Ly-$\alpha$ 
absorbers}

From the available literature we have collected data on LLS with $\tau_{LL}
>0.4$ and on DLS with rest frame equivalent width $W\ge 5$\AA\ for $z<4.7$. 
We have 661 QSOs and for each QSO we have the redshift path covered with a 
given sensitivity, the
redshift of the intervening absorbers and their estimated HI column density,
$N_{HI}$. We disregard absorbers and paths within 5000 km/s 
from the quasar redshift. The sensitivity of a search is expressed as a 
lower limit to the HI column densities 
searched. To a given direction it may correspond more than one path 
if different redshift paths were searched with different sensitivities.
If no estimates of the absorbers HI column density are
available (if no residual flux is detected  short ward of the Lyman break for
example) a lower limit to it is given while upper limits may be derived
from searches of damped Lyman-$\alpha$ absorption lines. The
HI column density values obtained from Voigt profile fits, when these are 
available, are included in our database. A comparison of the HI column 
densities
derived directly from the equivalent width, with those determined from 
the Voigt profile fit to the same absorption lines shows that the Voigt
fit procedure gives a systematically higher value of $N_{HI}$. This might be
due to an underestimate of the equivalent width due to an effective absorption 
of the quasar continuum radiation by intervening forest clouds. 
Using data where
Voigt fits are available we derive an average correction for $N_{HI}(W)$,
the column density derived from $W$, and we correct
the data whenever no Voigt fits are available as follows:

\begin{equation}
{\hbox{log}} N_{HI} = 6.261 +  0.705\ {\hbox{log}}N_{HI}(W), 
\label{ncorr}
\end{equation}

Since the maximum dispersion found in the data used for deriving eq. 
(\ref{ncorr}) is $\pm0.4$ in log$N_{HI}$ we use this value for  
estimating the errors on log$N_{HI}$ whenever we apply the above correction.
Our compilation of data is described in more detail in Paper II and is
available upon request from the authors.

\begin{figure}
\epsscale{.8}
\plotone{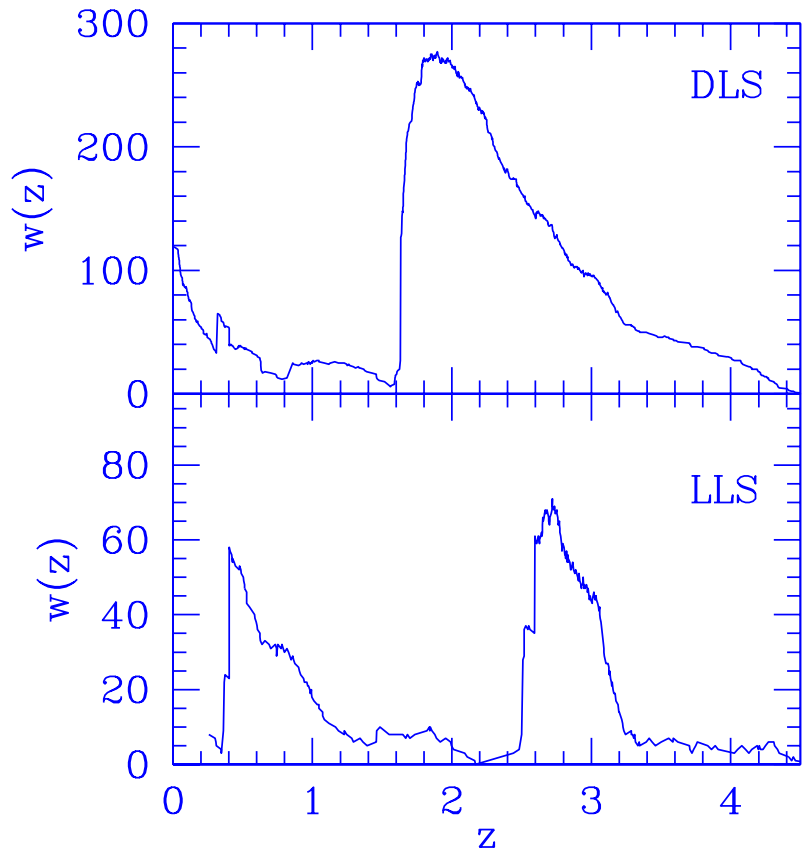}
\caption{The redshift path density covered by our collection of data.
 \label{fig1}}
\end{figure}

\placefigure{fig1}

The redshift path density $w(z)$ \citep{lwt91}, defined as the number 
of lines of sight at each redshift searched for LLS with any sensitivity 
or for DLS with $W\ge 5$, is shown in Figure 1 for all our data.
From our data set we extract the LLS sample and the DLS sample. 
The LLS sample consists of all observations which were sensitive 
to LLS with $\tau_{LL} \ge 1$. The DLS sample consists of observations 
sensitive to absorbers with column density log$N_{HI} \ge 20.13$.
We wish to examine the number density of absorbers per unit
redshift, $dN/dz$  in the LLS and DLS sample.
As usual we represent the number density as a power law of the 
form

\begin{equation}
N(z)= N_0 (1+z)^\gamma 
\end{equation}

In order to avoid binning the data in redshift intervals, which can 
generate errors in the estimate of the evolutionary trend, we use
the maximum likelihood method described by  \citet{sto94} to estimate 
$N_0$ and $\gamma$. 
For the LLS sample the maximum value of the Likelihood is found  
for:

\begin{equation}
\gamma = 1.66\pm 0.3 \qquad  N_0 = 0.23 
\label{gamlls}
\end{equation}

\noindent
where the errors quoted are the $> 68.3 \%$ confidence limits.
The $> 99.7 \%$ confidence limits for $\gamma$ are 0.80 and 2.64.
These values are similar to the ones obtained by \citet{ste95}, 
For the DLS sample we find 

\begin{equation}
\gamma = 0.86\pm 0.3 \qquad  N_0 = 0.18 
\label{gamdls}
\end{equation}

\noindent
The $> 99.7\%$ confidence limits are 0.11 and 1.72. The value of
$\gamma$ is consistent but slightly lower than what has been
found by \citet{sw00} for higher column density systems. If we
restrict our DLS sample to only those systems with have
a measured log$N_{HI} \ge 20.3$ we obtain the same values 
of $N_0$ and $\gamma$ as given by \citet{sw00}. 
Figure 2 shows the cumulative number of absorbers versus $z$ for
both samples. Over plotted is the expected number of LLS and DLS 
from the Maximum Likelihood estimate. Notice that 
DLS at $z\ge 1.75$ seem to require a lower value of $\gamma$
than what is given by the Maximum Likelihood estimate over the 
whole $z$ range. In fact by considering only redshifts $z\ge 1.75$
the match between the data and the best fit cumulative function 
for DLS improves: in this case $\gamma=-0.07$ with
$-1.58$ and 1.36 as $> 99.7\%$ confidence limits.

\begin{figure}
\epsscale{.8}
\plotone{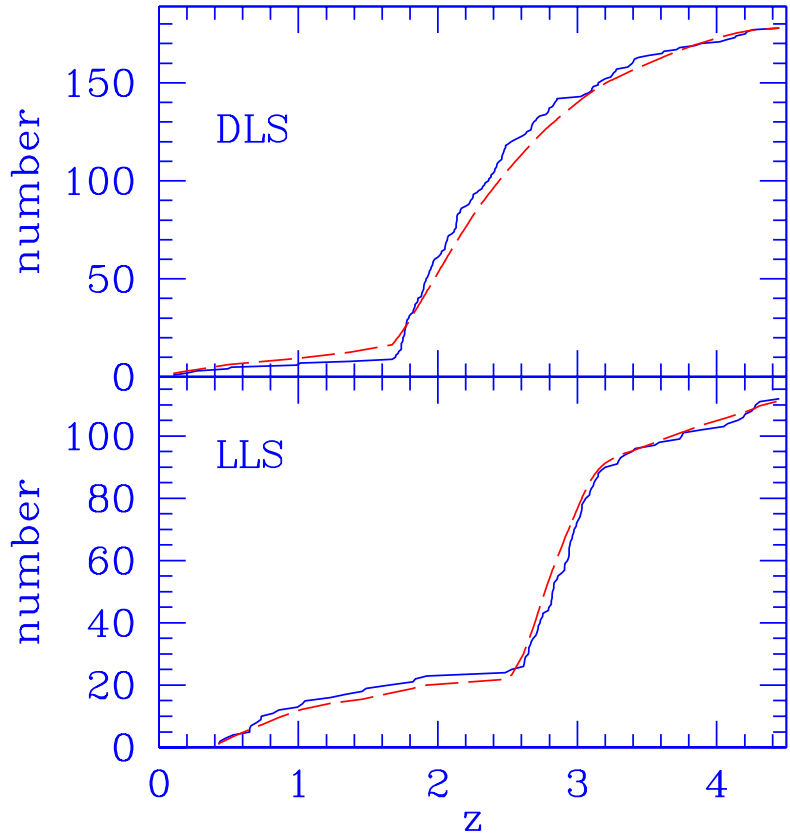}
\caption{The best fit to the cumulative 
LLS and DLS number evolution. \label{fig2}}
\end{figure}

\placefigure{fig2}

For the present paper we shall use data only in the intermediate 
redshift range  $1.75\le z <3.25$ and for this range  the
slight difference in evolution implied by eq. ~(\ref{gamlls})
and (\ref{gamdls}) is  unimportant. We  use
an average $\gamma$ value of 1.0 which is consistent both with 
the number density evolution of DLS and of LLS.
We exclude from the detections
absorbers which were declared as ``non damped'' if they were not
detected as LLS or for which no LLS searches have been performed. For
``non damped'' Ly-$\alpha$ absorbers which were detected as LLS
the $N_{HI}$ is set to the value derived from $\tau_{LL}$ when this is 
available; otherwise we shall use the lower limit to $\tau_{LL}$ for 
a lower limit to $N_{HI}$ and the column density value 
derived from $W$ as upper limit.

\section {The $N_{HI\perp}-N_{H\perp}$ curve}

Our numerical code solves for the ionization, chemical, and
hydrostatic equilibrium of a gaseous absorber  which
receives photons from a background radiation field.
It computes the HI column density perpendicular to the
plane, $N_{HI\perp}$, in a plane parallel geometry when the 
total hydrogen column density perpendicular to the plane is
$N_{H\perp}$ and the density stratification in the vertical 
direction $y$ is in the following exponential form:

\begin{equation}
\rho(y)=\rho_0 {\hbox{exp}}\Bigr(-{y^2 \over 2h^2}\Bigl)
\label{density}
\end{equation}

\noindent
$\rho_0= M_g/(\sqrt{2\pi}h)$ is the central gas 
density for a given total mass surface density 
per cm$^{-2}$ of the gaseous component, $M_g$.   
eq. (\ref{density}) is the exact solution for the Poisson and first
moment equation for an isothermal gaseous disk in equilibrium in a 
dark matter 
halo potential in the limit of a negligible self gravity and of 
negligible changes of the dark matter density
with height above the
midplane (within the gaseous extent, e.g. Spitzer 1942,
Maloney 1993). In this case the scale height can be written as

\begin{equation}
{1\over h^2}={4\pi G\rho_{dm} \over c_s^2}={(2\pi G)^2 M_{dm}^2
\over c_s^4}
\label{sigma1}
\end{equation}

\noindent
$c_s$ is the sound speed (computed using $T$, the mass averaged 
temperature of the slab), $\rho_{dm}$ and $M_{dm}$ are respectively
the dark matter volume density and the dark matter surface
density between +$h$ and -$h$ in the slab. 
In the absence of dark matter instead eq. (\ref{density})
is only an approximation to the self gravitating isothermal
gas layer solution (e.g. Spitzer 1942) 

\begin{equation}
\rho(y)=\rho_0 {\hbox{sech}}^2\Bigr({y\over \sqrt{2} h_{sg}}\Bigl)
\qquad {1\over h_{sg}^2}={2\pi^2 G^2 M_g^2 \over c_s^4}
\label{sigma_sg}
\end{equation}

\noindent
In this case the gas central density is $\rho_0=M_g/(2\sqrt{2} h_{sg})$.
If we approximate the above exact solution 
with eq. (\ref{density}), given the same 
central volume density $\rho_0$ and surface density $M_g$, we should 
use the following expression for the vertical scale height

\begin{equation}
{1\over h^2}= {\pi \over 4 h_{sg}^2} \simeq {16 G^2 M_g^2 \over c_s^4}
\label{sigma2}
\end{equation}

It will be shown in a subsequent section, that eq. (\ref{density}), 
with the scale height $h$ given by eq. (\ref{sigma2}), is a good
approximation to the exact self gravitating gas layer solution
for the purposes of this paper.
We can therefore write a general expression for $h$ if we use eq.
(\ref{density}) to describe the vertical stratification of the gas
when both self gravity and dark matter are taken into account:

\begin{equation}
{1\over h^2}={16 G^2 M_g^2\over c_s^4} (1 + \eta^2) \qquad
\qquad {\hbox{with}}\ \eta\equiv {\pi M_{dm}\over 2 M_g}
\label{sigma} 
\end{equation}

\noindent
The compression factor $\eta$ for the volume density is in 
general considered as the effect of gravity from matter other 
than the gas (dark matter, stars or brown dwarfs which coincide 
with the gas location). We write $\eta$ as 

\begin{equation}
\eta\equiv\eta_0 {c_s \over \tilde c_s }
\label{eta}
\end{equation}
 
\noindent
$\tilde c_s$ being the value
of $c_s$ for $T=10^4$ K and for a mean gas mass per particle 
$\mu=1.3$. In this paper we shall consider a set of models, each 
corresponding to a different $\eta_0$ value, and in each model
we keep $\eta_0$ constant as we very $M_g$ while $c_s$ is computed
from the radiative transfer and energy equation. A constant $\eta_0$
value for $\eta_0\gg 1$ implies  

\begin{equation}
\rho_{dm} \propto M_g^2
\label{eta0}
\end{equation}

\noindent
The meaning of the above proportionality are not immediate, but
for spherical isothermal dark matter halos eq. (\ref{eta0}) implies
a constant ratio between the gaseous and dark total surface 
densities. $\eta$ instead is
proportional to the the dark matter surface density between
$+h$ and $-h$ and since $h$ scales linearly with $c_s$, $\eta$
will also scale with $c_s$.

The equilibrium fractional ionizations of H,He,HeII  
are found step by step through the slab taking into
account the diffusion of photons from the recombination
of H, He, HeII and the secondary electrons produced
by the harder photons.
The temperature at each step through the slab is given by balancing
photoionization heating with 
the cooling from hydrogen and helium gas (collisional ionizations,
recombinations, lines excitation and free-free processes), 
from metal lines and from $H_2$ cooling. After each iteration
the mass averaged temperature $T$ is computed consistently and is used 
to determine the gas vertical dispersion and stratification. 
The ionization states of metals (C,O,Fe) are computed step by step
consistently with the photoionization rates and charge exchange reactions; 
very high ionization states (i.e. potential energies $E \gg 54$ eV)
are not considered. For metal abundance
we use $Z=0.02 Z_\odot$ throughout this paper but we
shall discuss briefly its possible variations.
H$_2$ fractions, although small, are computed  
via primordial chemical reactions \citep{cgp97}.

From the ionization-recombination balance we know that for a small
neutral fraction the ratio $N_{H\perp}/N_{HI\perp}$ depends  
on the ratio between the ionization and recombination coefficients
and varies inversely with the gas volume density. We define $X$ as  
 
\begin{equation}
X\equiv {\hbox{log}}{N_{H\perp}\over N_{HI\perp}}\qquad {\hbox{for}}
\ \ N_{HI}=1.6\times 10^{17} {\hbox{cm}}^{-2}
\end{equation}

\noindent
Notice that $X$ is defined at the line of sight value of 
$N_{HI}=1.6\times 10^{17} {\hbox{cm}}^{-2}$ and therefore the
corresponding $N_{HI\perp}$ value depends on the thickness ratio
$h/R$. We derive  for $\eta_0\ge 1$ the following relation

\begin{equation}
X\simeq 2.94 + 0.36\ {\hbox{log}}{J_L\over \eta_0}
\label{X}
\end{equation}

where $J_L$ is the intensity of the background flux at 912 \AA\  
written in
units of $10^{-22}$ ergs cm$^{-2}$ s$^{-1}$ Hz$^{-1}$ sr$^{-1}$.
For our redshift range ($1.75<z<3.25$) we use
a constant $J_L$ and a flux spectrum computed for $z=2.5$ 
by Haardt $\&$ Madau (1996, hereafter H-M) for emission by quasars and 
absorption by intervening clouds. A nearly constant ionization rate
between $z=2$ and 4 is also required by comparison of results from 
hydrodynamical simulations of structure formation with the measured
opacity of the Lyman-$\alpha$ forest. The H-M flux intensity at 
$z\simeq 2.5$, $J_L\simeq 5.5$,
agrees sufficiently well with other estimates \citep{rau97,gia96}
and we shall use $J_L =5.5$ throughout. The compression 
factor $\eta_0$ is unknown a priori and  will be determined by trial
and error. Hence $\eta_0$ (or equivalently $X$), and $\alpha$ are 
parameters to be determined. 
 
\begin{figure}
\epsscale{.99}
\plotone{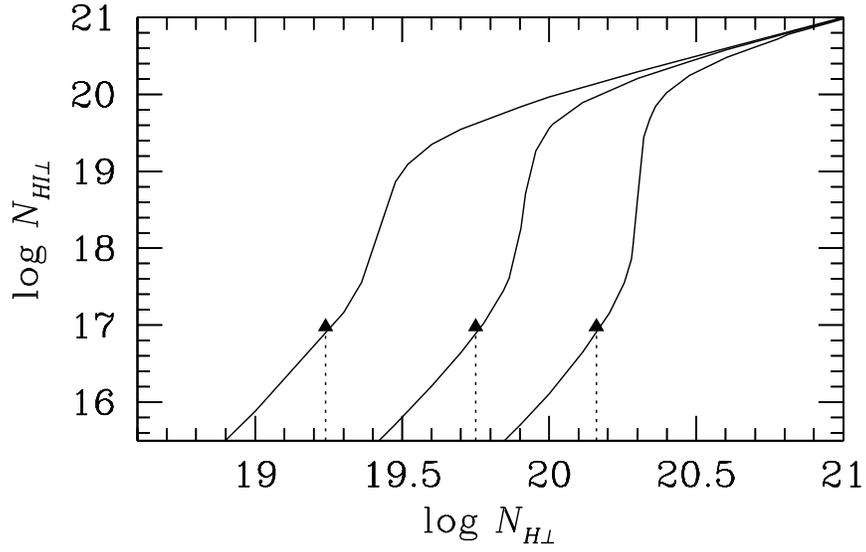}
\caption{
The log value of $N_{HI\perp}$ as a function of $N_{H\perp}$ 
for slabs irradiated by the H-M flux at  $z=2.5$ and for 
log${J_L/\eta_0}=-1.9,-0.4,0.7$ (from left to right). 
The dotted vertical lines indicate 
$N_{H\perp}=9.5\times 10^{16}$ cm$^{-2}$
corresponding to $N_{HI}=1.6\times 10^{17}$ cm$^{-2}$ for $h/R=0.2$.
\label{fig3}}
\end{figure}

\placefigure{fig3}

In Figure 3 we show three curves for $N_{HI\perp}$ as a function of 
$N_{H\perp}$ for log$J_L/\eta_0$
=$-1.9,-0.4,0.7$, which correspond to $X = $ 2.3, 2.8, and 3.2 respectively. 
All curves in Figure 3 are characterized by 3 regions: a first region
at high column density where $N_{H\perp}\simeq N_{HI\perp}$, a second region where
for a small change in $N_{H\perp}$ the HI column density changes very rapidly,
and a third region, towards the bottom part of the plot, where the log$N_{H\perp}$-log$N_{HI\perp}$ relation is again linear with a slope 
smaller than unity and which in general depends on the
variations of $\eta$ with $N_{H\perp}$. For the models in this paper, we
assume that $\eta_0$ in eq. (\ref{sigma}) does not vary with gas surface  
density $M_g$ (see section 3). The factor ${J_L/(1+\eta_0)}$ is what
determines the critical column density where the ${HI\rightarrow
HII}$ transition occurs i.e. the column density  where the 
steep slope of the second region starts. The
rapidity of the decrease of $N_{HI\perp}$ in the second region depends
also on metal abundances. When $Z$ increases above 
0.05 $Z_{\odot}$
metal line cooling becomes quite important in that region and 
the transition gets less sharp as $Z$ increases.
The shape of these curves is quite 
important because it determines the slope of the distribution function 
around the Lyman-Limit region.

\section {From the observed $N_{HI}$ distribution function to
the total amount of gas distributed in LLS and DLS.}

Several papers on the HI column density
distribution of absorbers \citep{tyt87,lwt91,pet93} have shown
that a power law  $N_{HI}^{-1.5}$ fits the data 
from $10^{13}$ cm$^{-2}$ to $10^{21}$ cm$^{-2}$ approximately, but 
not well enough
to satisfy statistical tests such as the Kolmogorov-Smirnov test. 
In this paper we shall use the deviations of $N_{HI}$ distribution
from a power law to determine $X$, after assuming that the total gas
column density $N_H$ has a power law distribution function of the form 

\begin{equation}
g(N_{H\perp})=
K (1+z)^{\gamma} N_{H\perp}^{-\alpha} 
\end{equation}

\noindent
In order to derive the $N_{HI}$ distribution
from $g(N_{H\perp})$ we must orient the absorbers randomly in the plane  
of the sky, assume an axial ratio $h/R$ for the slab, and apply the 
$N_{H\perp}-N_{HI\perp}$ conversion factor. For our model fitting we 
shall use $Z=0.02
Z_\odot$,  $X$ independent of redshift and an axial ratio of $h/R=0.2$, since
clouds are likely to be neither spherical nor thin disks. Results essentially 
do not depend on $h/R$ for $h/R\le 0.5$.
We derive $\alpha$ and $X$ by comparing the resulting $f(N_{HI})$
for $\gamma=1.0$ with the data present in our compilation at 
$N_{HI}\ge 1.6\times 10^{17}$ cm$^{-2}$. $K$ is set by the normalization
condition for $g$, based on the observed number of absorbers.

We emphasize that it is not possible to present the data
relative to the distribution function in a model independent way, 
due to LLS with undetermined $\tau_{LL}$.  A deterioration of the 
available data might result from the operation
of binning in $N_{HI}$ in order to render straightforward the
comparison with the model distributions. Instead of binning the data, we
match the model distribution to the individual detections. 
The details of the fitting procedure are given in Paper II; we underline 
here the main characteristics.
The procedure is rather similar to that used by \citet{sto96}, but 
implementing the algorithm in order to
take into consideration also the uncertainty in the determination
of any single value of $N_{HI}$. Large observational errors are
included by leaving undetermined the ``real'' position of each event.
We determine $\alpha$ and $X$ by a Maximum Likelihood 
analysis to the projected HI column density distribution $f$ fixing
the real position of each event to the measured value of $N_{HI}$
when this is available; otherwise we use in the Likelihood the
integral of $f$ between the  maximum and minimum value of $N_{HI}$.
We normalize the theoretical distribution such as to give a number of 
detections with $N_{HI}\ge 1.6\times 10^{17}$ cm$^{-2}$ equal to the
observed one. Two maxima for the Likelihood are found:

\begin{equation}
X=2.82 \qquad \alpha=2.70 \qquad K=1.2\times 10^{34}
\label{bestfit1}
\end{equation}

\begin{equation}
X=2.74 \qquad \alpha=2.57 \qquad K=1.8\times 10^{31}
\label{bestfit2}
\end{equation}

The $>68.3\%$, $>95.5\%$, and $>99\%$ confidence levels in the $X-\alpha$
plane are shown in Figure 4 where the filled dots indicates the location
of the maxima as in eq. (\ref{bestfit1}) and (\ref{bestfit2}).
The self gravitating gas solution ($\eta=0$, $\alpha=4.63$, $X=3.33$) lies 
well outside the $>99\%$ confidence level (we would need the 99.999$\%$
confidence level to include it) and therefore it is not
consistent with the data. We have also checked that a similar
conclusion holds if we use the exact self gravitating
gas solution, as given by eq. (\ref{sigma_sg}), in deriving
the $N_{HI\perp}-N_{H\perp}$ relation. For this case both 
$X$ and the best fit $\alpha$ value are within $1\%$ of the values
obtained using eq. (\ref{density}) and $\eta=0$ 
for the vertical gas stratification.

\begin{figure}
\epsscale{.99}
\plotone{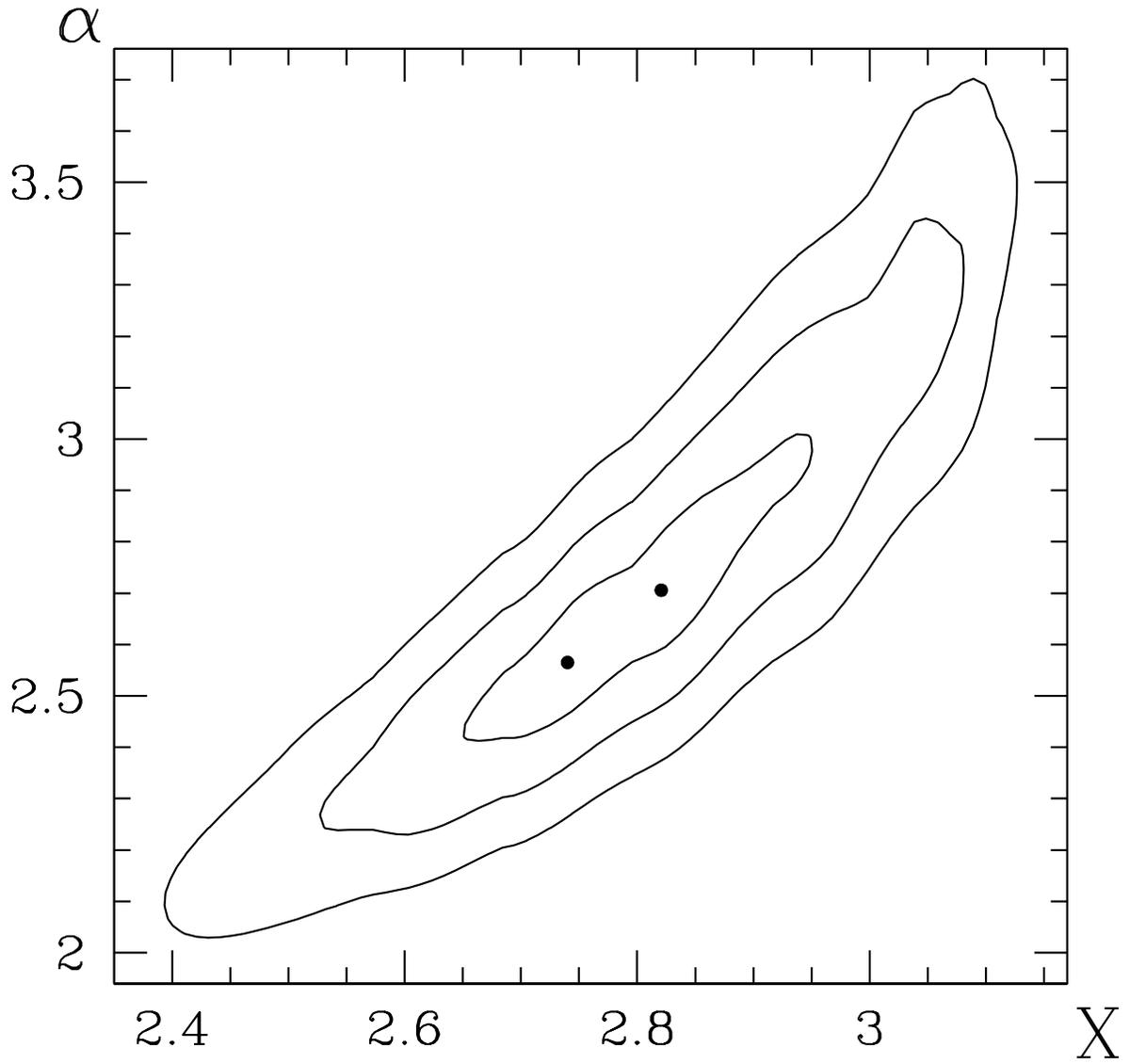}
\caption{The $>68.3\%$, $>95.5\%$, and $>99\%$ confidence levels in the 
$X-\alpha$ plane.
\label{fig4}}
\end{figure}

\placefigure{fig4}

In Figure 5 we compare the observed value of the cumulative function 
with the theoretical ones derived from the integral of the projected 
HI column density; we show results for the two best fit models (the two
maxima in Figure 4) and for two models corresponding to the highest and lowest 
X values on the $>95.5\%$ confidence level of Figure 4. 
For points which have no defined $N_{HI}$, i.e. which have large errors,
we then compute their best 
distribution for a given $f$ by spreading the data in the
allowed range of $N_{HI}$ according to $f$ weighted with the 
redshift path. Between all the
possible permutations of points with undetermined $N_{HI}$ we then 
choose those which satisfies best the $U$-test on the deviations
between the observed and the expected cumulative functions over the error interval, 
$R$, and over $N_{HI}$, $C$ (see Paper II for more details
and for the use of a numerical simulation to proof the validity of
this approach).
For $X-\alpha$ inside the $99\%$ confidence level of Figure 4
the K-S tests on $R$ and $C$ are satisfied to the 99.9$\%$ level. 

\begin{figure}
\epsscale{.99}
\plotone{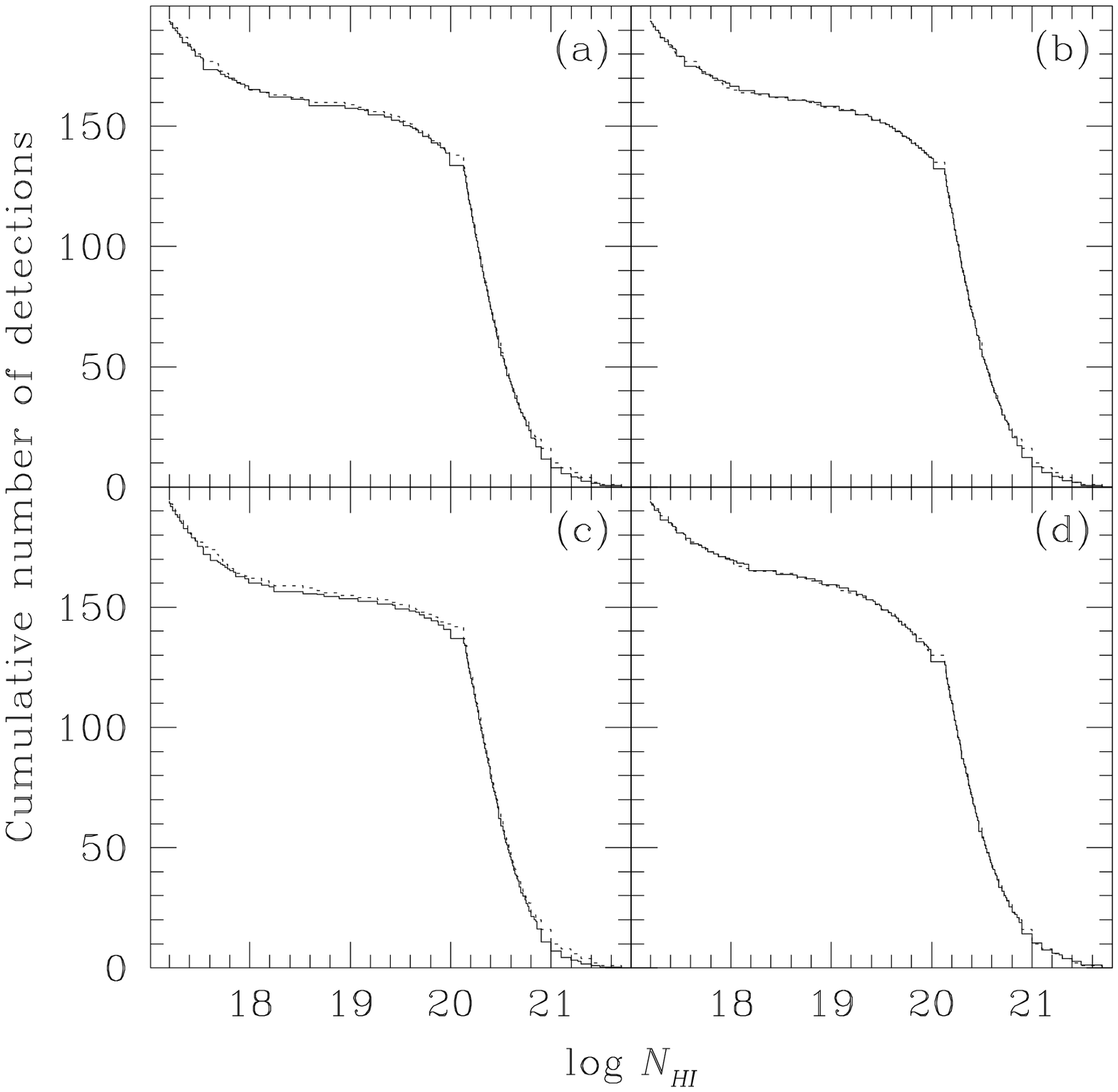}
\caption{The theoretical cumulative distribution functions (continuous lines) 
for our two best fit models ($(a)$ $\alpha=2.70$, $X=2.82$ and 
$(b)$ $\alpha=2.57$, $X=2.74$)
and for two points over the $>95.5\%$ confidence level 
($(c)$ $\alpha=3.32$, $X=3.07$ and $(d)$ $\alpha=2.27$, $X=2.53$). For comparison 
the observed HI cumulative functions (dotted lines) are shown.
\label{fig5}}
\end{figure}

\placefigure{fig5}

We have proved that the gas distribution between the LLS and the DLS 
region follows a single power law with index $\alpha > 2$ if the
ionization level is such that less than $1\%$ of the total gas
is neutral when $N_{HI}=1.6 \times 10^{17}$ cm$^{-2}$.
There is no need of a distribution more complicated than a power law
once one takes into account ionization effects.
Our results on $\alpha$ and $X$ still hold 
even  if we do not include in the data set the damped lines with
$5\le W\le 10$ \AA\ or if we exclude a certain percentage of these
lines due to possible blending with smaller lines.
 
\subsection{The total gas content of LLS and DLS}

We shall discuss here some results relative to 
the best-fit values of $X$ and $\alpha$ as given in eq. (\ref{bestfit1})
and in eq. (\ref{bestfit2}), 
and to the two most extreme values of $X$ on the $>95.5\%$ confidence level, 
namely:  $X=3.07$ ($\eta_0=2.3$), $\alpha=3.32$ (hereafter $2\sigma$-up);
$X=2.53$ ($\eta_0=75$), $\alpha=2.27$ (hereafter $2\sigma-$low).
The distribution function for the total column density $\tilde 
N_{H\perp}\equiv N_{H\perp}/10^{20}$ cm$^{-2}$ can be written as 

\begin{equation}
g(\tilde N_{H\perp})=
\tilde K (1+z) \tilde N_{H\perp}^{-\alpha} \qquad
\label{gnorm}
\end{equation}

In Figure 6, in arbitrary scale, the continuous 
lines show log $f(N_{HI})$, the HI
distribution function,  for the best-fit values of $X$ and $\alpha$ as given in 
eq. (\ref{bestfit1}) and (\ref{bestfit2}), and for the $2\sigma-$low and
the $2\sigma$-up models. 
Our data for $N_{HI}>1.6\times 10^{17}$ cm$^{-2}$ is in five large bins just
for the purpose of presentation. 

\begin{figure}
\epsscale{.99}
\plotone{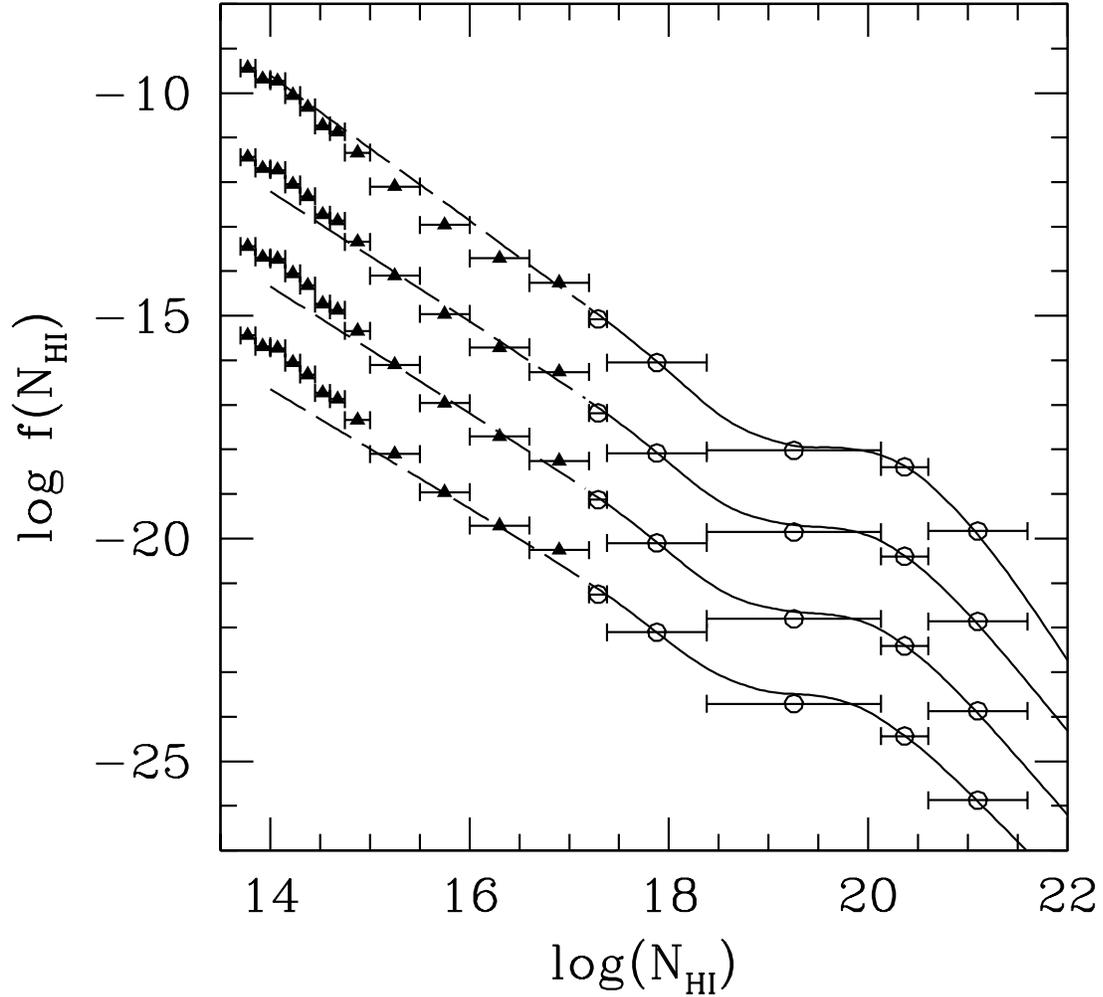}
\caption{HI column density distribution functions between $z=1.75$ and
$z=3.25$ for $X=3.07,2.82,2.74,2.53$ and $\alpha=3.32,2.70,2.57,2.27$ respectively 
(from top to bottom). Bins with open circles have data from our 
collection while filled triangles are data from the compilation of 
\citet{pet93}. Scale is arbitrary for each theoretical model but  
data points have been scaled accordingly.The long dashed line indicates the 
extrapolation of the distribution function towards the forest region
using the same $\eta_0$ value as for the LLS and DLS. 
\label{fig6}}
\end{figure}

\placefigure{fig6}

$\tilde N_{H\perp}\times g(\tilde N_{H\perp})$ can be integrated over a 
range of $\tilde N_{H\perp}$, say between  $\tilde N_{H\perp,i}$ and 
$\tilde N_{H\perp,n}$,
to estimate the mass density of hydrogen atoms in gas clouds with 
an average HI column density along the line of sight between 
$N_{HI,i}$ and $N_{HI,n}$. The comoving 
cosmological H+He gas density at $z=2.5$ can be written as:

\begin{equation}
{\Omega_{gas}(i)h_{60}} \simeq 
     {2\times 10^{-3} \tilde K \over 2-\alpha}
     {(1+z)\over (1+z)^\delta}
     ( \tilde N_{H\perp,i}^{-\alpha+2}- \tilde N_{H\perp,n}^{-\alpha+2}) 
\label{omega}
\end{equation}
 
\noindent
$\delta$ depends on the cosmological model and is a function of $z,\Omega_M,
\Omega_\Lambda$. For a standard Friedmann
Universe in which $q_0=0$, $\delta=1$ and we shall use this value for the
rest of this section (for $\Omega_M=0.3$ and $\Omega_\Lambda=0.7$ instead
$\delta$ depends on $z$ and is close to zero at $z\sim 2.5$). 

In Table 1 we give the values of 
$\Omega_{gas}(i)h_{60}$, for $i=1,2,3$ which is the total gas density
in the Universe at $z\simeq 2.5$ due to absorbing
clouds whose HI column density projected along the line of sight 
is between $N_{HI,i}$  and $10^{22}$ cm$^{-2}$. We shall consider 
$N_{HI,1}=10^{14}$ (see section 4.2), 
$N_{HI,2}=1.6\times 10^{17}$ and $N_{HI,3}=1.3\times 10^{20}$ cm$^{-2}$. 
For each corresponding value of $N_{H\perp,i}$ we give $X_i$, the log ratio 
of total to neutral gas column density. Results are given both for the
best fitting models and for the two most extreme values of 
$X$ on the $>95.5\%$ confidence level of Figure 4. In the Table
we also show the gas scale heights for 
$N_{HI,2}=1.6\times 10^{17}$ and $N_{HI,3}=1.3\times 10^{20}$ cm$^{-2}$,
and values of $\Omega_{dark}(i)$. 

For $\Omega_{dark}$ in the regions coinciding with the gas the factor
$\zeta$ to be substituted into Table 1 is $\zeta\approx 1$, independent
of assumptions on cloud size and relative distribution of dark matter
and gas. For rotating disks embedded in  spherical dark halos
one can compute the contribution of the total dark matter surface
density to the cosmological matter density. This contribution
associated with DLS or LLS systems depends on the rotational
velocity $V$ and is given by $\Omega_{dark}$
using $\zeta \approx V/\tilde c_s$. This
factor may be close to one for dwarf clouds, but $\zeta\gg 1$ would hold
for giant disk proto-galaxies.
For the range of uncertainties in Table 1, the value of $\Omega_{gas}(2)$,
the total contribution of LLS plus DLS, varies little but $\eta_0$ has a
large spread. Consequently, $\Omega_{dark}$ and the gas scale height $h$
also have a large spread. Note that $h(3)$ is particularly small for
the $2\sigma$-low limit.

\subsection{Conjectures on the Lyman-$\alpha$ forest}

The Lyman-$\alpha$ forest clouds may not be directly related to LLS, but we 
can explore the consequences of two assumption, namely that eq. (\ref{gnorm})
extends into the Lyman-$\alpha$ forest region with a compression factor
$\eta_0$ in eq. (\ref{sigma}) that has the same constant value as for LLS and DLS.
The filled triangles in Figure 6 are the observed data for the Lyman-$\alpha$
forest from \citet{pet93} and \citet{hu95}. The dashed curves show the extrapolations
of the distribution functions keeping $\alpha$ and $\eta_0$ constant.
Note that these extrapolations, without any additional parameters, give
surprisingly good fits, especially the best fit model given by eq. (\ref{bestfit1})
and the $2\sigma$-up model
(the $2\sigma-$low model and the best fit model given by eq. (\ref{bestfit2})
favour a different mechanism of confinement in place for the
very low column density clouds, like
external pressure, which would steepen the distribution by keeping the fractional 
ionization of H independent from the cloud column density). 

If Lyman-$\alpha$ forest clouds were physically different from LLS and DLS,
one might have expected a change in $\alpha$ or $\eta_0$ and hence a much worse
fit of the dashed curves in Figure 6. The good fit makes it even more likely that
$\alpha$ and $\eta_0$ are constant from LLS to DLS, which represents a much
smaller range in $N_H$.

\section {Summary and Discussion}

The results of the model fit to LLS and DLS presented in this paper
underline new aspects in the column density distribution 
of Lyman-$\alpha$ absorbers:

$\bullet$ The excess of systems in the damped region, suggested by the 
$N_{HI}$ data, is naturally explained in terms of a sharp
transition from a highly ionized to a highly neutral gas
distribution and it does not requires any break in the distribution of the 
total gas column density.  

$\bullet$ The total gas column density distributions, $g(N_{H\perp})$, which 
best fits the data for $1.75\le z <3.25$ in the LLS and DLS region can be
described by a power law of index $-\alpha$ with $2\le \alpha \le 3.7$.
This has the important consequence that low column density systems
contains more mass than high column density systems.

$\bullet$ We have tested that our results, which are relative to the
redshift bin $1.75\le z \le 3.25$, depend weakly on the data
selection and on the number redshift evolution. They also do not show a 
strong dependence on the assumed thickness to diameter ratio of the slab 
or from its metallicity if $Z\le 0.05 Z_\odot$.

$\bullet$ The gas fractional ionizations increase with  decreasing column 
density and data are best fitted when hydrogen fractional ionizations 
are of order $\sim 0.002$ at $\tau_{LL}=1$ i.e. $X\sim 2.8$. Gas fractional 
ionizations for absorbers in a background radiation field depends on the 
gas volume densities and temperature.
The model presented in detail in this paper considers the gas self gravity 
and a dark matter potential in which the dark matter surface
density inside one gas scale height is proportional to the gas surface 
density (constant $\eta_0$). However the range of 
$\alpha$ and $X$ values compatible with the data are similar if one considers 
instead a model where $\eta_0$ varies with $N_H$ (e.g. if $\eta_0 \times N_H$
is kept constant instead). Table 1 summarizes our main results; the values
of $\eta_0$ for more general models refer to LLS column densities which are no
longer opaque to the UV ionizing radiation.

Although our Likelihood analysis allows power law indices $\alpha$ in
a rather wide range, 2 to 3.7, other arguments make the upper half of
this range more likely: $\alpha \ge 2.7$, $\eta_0 \le 13$, $X \ge 2.8$.
For larger values of $\eta_0$ the gas scale height $h(3)$ for DLS is
uncomfortably small and the LLS contribution $\Omega_{dark}(2)$ to
cosmological matter density is uncomfortably large (see Table 1, the 
$\Omega$ values given are for $q_0=0$; $\Omega$ values for $\Omega_M\sim 0.3$
and $\Omega_\Lambda=0.7$ are larger by a factor $\sim 2.5$). For the total
dark matter mass, the factor $\zeta$ for $\Omega_{dark}$ is of order 
$V/10$ km s$^{-1}$, which suggests $V<100$ km s$^{-1}$ on average for 
LLS and DLS at redshifts $z\sim 2.5$.
This finding is in agreement with the rotational velocities of absorbers halos 
predicted by \citet{vss99} and with models that
propose that a large fraction of Lyman-$\alpha$ absorption systems originate
from small low luminosity systems \citep{am98,rt00,hsr00}.
However \citet{sdp94} have shown that small rotational velocities are not 
required by low redshifts MgII absorption systems with $W>0.3$ \AA. 
These have the same average density per unit path of redshift as LLS and
seem to be associated with normal galaxies whose halo extends for $\sim 40$
kpc. This opens the issue about possible evolutionary scenarios
for LLS, which can be fully addressed only once we know 
the redshift evolution of galactic halos and of the galaxy 
luminosity function. The possible role of 
faint galaxies as those in pairs close to bright galaxies 
\citep{chu00} should also be taken into account. Futhermore it would be  extremely useful to have a larger statistical sample of LLS and DLS at 
$z<1.75$ and at $z>3.25$ which, together with a better
knowledge of the evolution of the background ionizing radiation field, 
allows a determination of $\alpha$ and $\eta$ at these redshifts
(see Paper II for an attempt with the actual data set).

$\bullet$ Although Lyman-$\alpha$ forest clouds, with their much smaller 
column density,
might be physically different and have different values of $\alpha$ and 
$\eta_0$, extrapolations using constant $\alpha$ and $\eta_0$ values
as for LLS and DLS give surprisingly good fits.
 
We cannot make definitive statements about the low density Lyman-$\alpha$
forest clouds because of possible dynamic deviations from hydrostatic
equilibrium and possible pressure confinement. Nevertheless, the fact that
$\alpha$ is appreciably larger than 2 near LLS for $z\sim 2.5$ suggests that
forest clouds contained more gas than LLS, which in turn contained more gas 
than DLS. Several papers have shown that the gas content in DLS at  
$z\sim 2.5$ is
below the mass content in galaxies in the local Universe \citep{sw00}, 
$\Omega_{s+g} h_{60}\simeq 3-7\times 10^{-3}$ and that the
stellar mass density in stellar systems at $z\sim 2.5$  is very small
\citep{mpd98}. It is therefore likely that gas
in forest clouds and LLS at $z\sim 2.5$ has  collapsed and contributed
to stars in present day galaxies.

We are grateful to P. Madau for providing us numerical results on
the background radiation field, to the referee and to Dr. Bothun for
very useful comments to the original manuscript, and to D. Chernoff
for helping with the Arcetri-Cornell connection. One of us (RB) 
acknowledges partial support by NSF-PHY94-07194.

\clearpage
\begin{center}
\begin{deluxetable}{ccccccc}
\footnotesize
\tablecaption{Ionization and gas content of Lyman-$\alpha$
absorption clouds. \label{tbl-1}}
\tablewidth{0pt}
\tablehead{
\colhead {} & \colhead {2$\sigma$-up} 
& \colhead {best fit 1} & \colhead {best fit 2} 
&\colhead {2$\sigma$-low}
} 
\startdata
{$\alpha$} & 3.32 & 2.70 & 2.57 & 2.27 \\ 
{$\tilde K$} & 3.81 & 0.97 & 0.74  & 0.42 \\ 
{$\eta_0$} & 2.3 & 12.5 & 20. & 75. \\ 
{$X(1)$} & 5.5 & 5.2 & 5.1  & 4.9 \\ 
{$X(2)\equiv X$} & 3.1 & 2.8 & 2.7 & 2.5 \\ 
{$X(3)$} & 0.4 & 0.2 & 0.2 & 0.1 \\ 
{$h(2)$/kpc} & 2.6 & 0.8  & 0.6 & 0.3 \\  
{$h(3)$/kpc} & 1.0 & 0.3 & 0.2 & 0.05 \\ 
{$\Omega_{gas}(1)h_{60}$} & 0.06 & 0.01  & 0.01 & 0.006 \\ 
{$\Omega_{gas}(2)h_{60}$} & 0.005 & 0.004 & 0.004 & 0.003\\
{$\Omega_{gas}(3)h_{60}$} & 0.002 & 0.002 & 0.002 & 0.002 \\ 
{$\Omega_{dark}(1)h_{60}$} & 0.1$\zeta$ & 0.2$\zeta$ & 0.2$\zeta$ & 0.5$\zeta$ \\
{$\Omega_{dark}(2)h_{60}$} & 0.01$\zeta$ & 0.05$\zeta$ & 0.07$\zeta$ & 0.3$\zeta$ \\
\enddata
 
\end{deluxetable}
\end{center}

\clearpage

\end{document}